\documentclass[conference]{IEEEtran}
\IEEEoverridecommandlockouts
\usepackage[numbers,sort&compress]{natbib}
\usepackage{amsmath,amssymb,amsfonts}
\usepackage{algorithmic}
\usepackage{graphicx}
\usepackage{textcomp}
\usepackage{xcolor}
\def\BibTeX{{\rm B\kern-.05em{\sc i\kern-.025em b}\kern-.08em
    T\kern-.1667em\lower.7ex\hbox{E}\kern-.125emX}}

\begin{document} 
 
\title{Explainable Lung Disease Classification from Chest X-Ray Images Utilizing Deep Learning and XAI\\
 }
 
\author{
\IEEEauthorblockN{Tanzina Taher Ifty*, Saleh Ahmed Shafin*, Shoeb Mohammad Shahriar, Tashfia Towhid \thanks{*First two authors contributed equally to this research.}} 
 Department of Computer Science and Engineering\\
 Ahsanullah University Of Science and Technology (AUST), Dhaka, Bangladesh\\
 
Email-  \{tanzina19taher,salehahmedshafin7,tashfia6288\}@gmail.com, shoeb.cse@aust.edu
}

%\IEEEauthorblockA{\textit{dept. name of the organization (of Aff.)} \\
%\textit{name of the organization (of Aff.)}\\
%\and
%\IEEEauthorblockA{\textit{dept. name of the organization (of Aff.)} \\
%\textit{name of the organization (of Aff.)}\\
%\and
%\IEEEauthorblockA{\textit{dept. name of the organization (of Aff.)} \\
%\textit{name of the organization (of Aff.)}\\
%\and
%\IEEEauthorblockA{\textit{dept. name of the organization (of Aff.)} \\
%\textit{name of the organization (of Aff.)}\\
%\and
%\and

\maketitle 
\begin{abstract}
Lung diseases remain a critical global health concern, and it's crucial to have accurate and quick ways to diagnose them. This work focuses on classifying different lung diseases into five groups: viral pneumonia, bacterial pneumonia, COVID, tuberculosis, and normal lungs. Employing advanced deep learning techniques, we explore a diverse range of models including CNN, hybrid models, ensembles, transformers, and Big Transfer. The research encompasses comprehensive methodologies such as hyperparameter tuning, stratified k-fold cross-validation, and transfer learning with fine-tuning.Remarkably, our findings reveal that the Xception model, fine-tuned through 5-fold cross-validation, achieves the highest accuracy of 96.21\%. This success shows that our methods work well in accurately identifying different lung diseases. The exploration of explainable artificial intelligence (XAI) methodologies further enhances our understanding of the decision-making processes employed by these models, contributing to increased trust in their clinical applications. 
\end{abstract}


\section{Introduction}

Lung diseases affect a notable portion of the worldwide population and are a serious health problem. Advance detection and pinpoint diagnosis of lung disorders are crucial for both better patient outcomes and successful therapy. Chest X-ray (CXR) is a more inexpensive substitute to techniques like Polymerase Chain Reaction (PCR) and Computed Tomography (CT) scans for the advanced diagnosis of lung diseases \cite{1}. Modern expansion in artificial intelligence (AI) and deep learning have made it attainable to automatically point out lung diseases from chest X-ray pictures, with uplifting outcomes. These AI algorithms' pace and efficacy are especially convenient in emergency scenarios, where they upgrade resource management and accelerate decision-making.

The World Health Organization (WHO) stated Covid-19 a pandemic, and it has had a destructive consequence on human beings all around the globe \cite{2}. Pneumonia was the peak cause of death for children under five in 2019, considering nearly a third of the demises in this age group, and led to 2.5 million deaths \cite{3}.
After COVID-19 and HIV/AIDS, tuberculosis is the second leading cause of mortality from contagious diseases worldwide, placing 13th among death causes \cite{4}. Advanced lung illness spotting is crucial such as a backdrop.
 Deep learning models are proficient in picking out complex patterns in X-ray images that might be missed by humans, thereby helping in the early diagnosis of diseases. With traditional diagnostic practices grappling to identify subtle patterns in X-ray images, this research focuses on ameliorating lung disease detection using advanced deep-learning techniques.

Deep learning has been extensively used in recent years for the automated classification of medical pictures. In this paper, we develop deep Learning (DL) and transformer-based classification models for the diagnosis of four different types of lung disorders - Bacterial Viral Pneumonia, Tuberculosis, and COVID-19 and as well as Normal lungs. We use pre-trained CNN models, hybrid models, ensemble models, transformer models, K-fold technique, and hyperparameter tuning to train and test our model. We used Lung Disease Dataset \cite{5} obtained from kaggle website. We also use Explainable Artificial Intelligence (XAI) \cite{6} techniques such as Gradient-weighted Class Activation Mapping (Grad-CAM) \cite{7} and Local Interpretable Model-Agnostic Explanations (LIME) \cite{8}  to provide visual explanations for the model's predictions. These techniques will enable us to identify the significant features and areas of the chest X-ray images that contribute to the classification decisions, which can be useful for clinicians in the diagnosis and treatment of lung diseases. 

Our research's contribution is: (1) it outperformed the baseline results in terms of multi-class classification scores, and (2) our work encompasses a comprehensive set of popular approaches, leveraging XAI.

\section{Related Works}

There is a lot of work on detecting disease from chest X-ray images \cite{9}. Many approaches have been done for
identifying the disease, we focused on ways related to our task that are different from deep learning methodology. 

In order to classify CXR pictures into three groups—COVID-19, respiratory infections, and regular classes—Mohammad Rahimzadeh et al. trained deep convolutional networks. Their average accuracy was 99.50\% using two open-source datasets: 180 COVID-19, 6054 respiratory infections, and 8851 regular pictures. In particular, sensitivity for COVID-19 was 80.53\%, which helped to achieve an accuracy of 91.4\% overall. They used the loss function named categorical cross entropy loss and optimizer as Adam for training a concatenation neural network. To improve training effectiveness, data augmentation approaches were used \cite{10}.

To overcome the limitations of conventional RT-PCR testing techniques, the research offers a deep learning-based approach for COVID-19 detection utilizing chest X-ray pictures. 99.7\%, 95.02\%, and 94.53\% testing accuracies are attained by using a CNN architectural model evolved in the study article for a 2 classes, 3 classes, and 4 classes classifications. However, more details on the dataset used its size, composition, and potential biases or limitations are needed. The paper claims superiority over other related works in the field but requires more in-depth comparisons and discussions. The high testing accuracy of the proposed model demonstrates its potential usefulness, but further exploration and analysis are needed to fully evaluate its performance compared to other existing approaches \cite{11}.

The aim of the research article "Coronet: A DL Network for Identification and Diagnosis of COVID-19 from Chest X Ray Images" is to give a thorough and conclusive diagnostic technique using CXR for detecting the COVID 19 virus. The paper shows CoroNet, a Convolutional Neural Network (CNN) model trained on a dataset consisting of COVID-19 acquired from broad sources and chest X-ray images from pneumonia cases. This model showed a high overall accuracy of 89.6\% with precision of 93.5\%, recall of 98.2\% for COVID-19 cases in the 4-class classification, and 95.5\% in the 3-class classification. The study highlighted how helpful CoroNet may be in aiding medical professionals and radiologists during the pandemic. However, it confesses the necessity for further training data and advocates doing more validations and assessments on outside datasets to verify CoroNet's reliability and applicability.
\cite{12}.

The work offers a practical method for detecting COVID-19 in areas with limited resources by using chest X-ray imaging. It reports the challenges of identifying COVID-19, specifically in regions without access to biotechnology testing. The advocated approach requires taking chest X-ray pictures, extracting important characteristics from them, and then utilizing machine learning classifiers to split the pictures into different sets. The best application combines an ensemble of subset discriminant classifiers with ResNet-50 for important feature (DF) computation. For the five-class classification task, the application gets a detection accuracy of 91.6\% and is computationally efficient. The dataset, which presents that the suggested pipeline can accurately categorize COVID-19 classes, is accessible for more research and comparisons.
 \cite{13}.

Multi-scale convolutional Neural Network (MS-CNN), a Deep Learning framework evolved by Ovi Sarkar et al., is effective in accurately identifying six lung-related disorders. Lung opacity, COVID-19, fibrosis, TB, viral pneumonia, and bacterial pneumonia are among these illnesses. They used explainable AI (XAI) to increase prediction ability, which improved the diagnostic threshold for lung diseases. The accuracy and understanding of the model were improved by integrating XAI techniques like Grad-CAM and SHAP. The MS-CNN model outperformed other models and showed remarkable efficiency in detecting COVID-19, with an amazing accuracy of 96.05\%. 
\cite{14}.

\section{Dataset Analysis}
We used a public dataset of chest X-ray pictures named the Lungs Disease Dataset (4 kinds) \cite{5}. The dataset is divided into three directories—test, train, and validation—and comprises 10095 photos. The Test folder has 2025 photos, the Train folder contains 6054 images, and the Val folder contains 2016 images. Table 1 displays the dataset's class distribution.

It is prepared from various datasets. These datasets include COVID-19 Detection X-Ray Dataset \cite{15}, Lungs Dataset \cite{16}, Chest X-Ray Images (Pneumonia) \cite{17}, Chest X-Ray (Pneumonia,Covid-19,Tuberculosis) \cite{18}, Chest X-Ray 14 Dataset with Lungs Cropped \cite{19} and Tuberculosis (TB) Chest X-ray Database \cite{20}. It is combined to remove the same images in the dataset using VisiPics  [When two identical pictures are stored in different formats or resolutions, Visipics will identify them as duplicates even if they are identical but for little aesthetic differences] \cite{21}. 

\begin{table}[htbp]
\caption{Data Distribution}
\begin{center}
\begin{tabular}{|p{1.2cm}|p{1.1cm}|p{1.1cm}|p{1cm}|p{1cm}|p{1cm}|}
\hline

        \hline
        \textbf{Set Name} & \multicolumn{5}{c|}{\textbf{Classes}} \\ 
        \hline
        & \textbf{Bacterial Pneumonia} & \textbf{Corona Virus}  & \textbf{Normal} & \textbf{Tuber-culosis} & \textbf{Viral Pneumonia} \\
        \hline
        Test & 403 & 407 & 404 & 408 & 403\\
        \hline
        Train & 1205 & 1218 & 1207 & 1220 & 1204 \\
        \hline
        Validation & 401 & 406 & 402 & 406 & 401 \\
        \hline

\end{tabular}
\label{tab1}
\end{center}
\end{table}

Finally, it can be said that the dataset is split into 20.06\% test data, 59.97\% of
train data, and 19.97\% of validation data. Our deep learning models for the identification and categorization of lung illnesses are developed and assessed using this dataset.

\begin{table}[htbp]
\caption{Dataset orientation.} 
\begin{tabular}{|p{0.2\linewidth}|p{0.2\linewidth}|p{0.2\linewidth}|p{0.2\linewidth}|}
\hline
& \textbf{Test Data} & \textbf{Train Data} & \textbf{Validation Data} \\
\hline
\textbf{Percentage} & 20.06\% & 59.97\% & 19.97\% \\
\hline
\end{tabular}
\end{table}

\section{Methodology}

\begin{figure}[htbp]
\centering
\includegraphics[width=1\hsize]{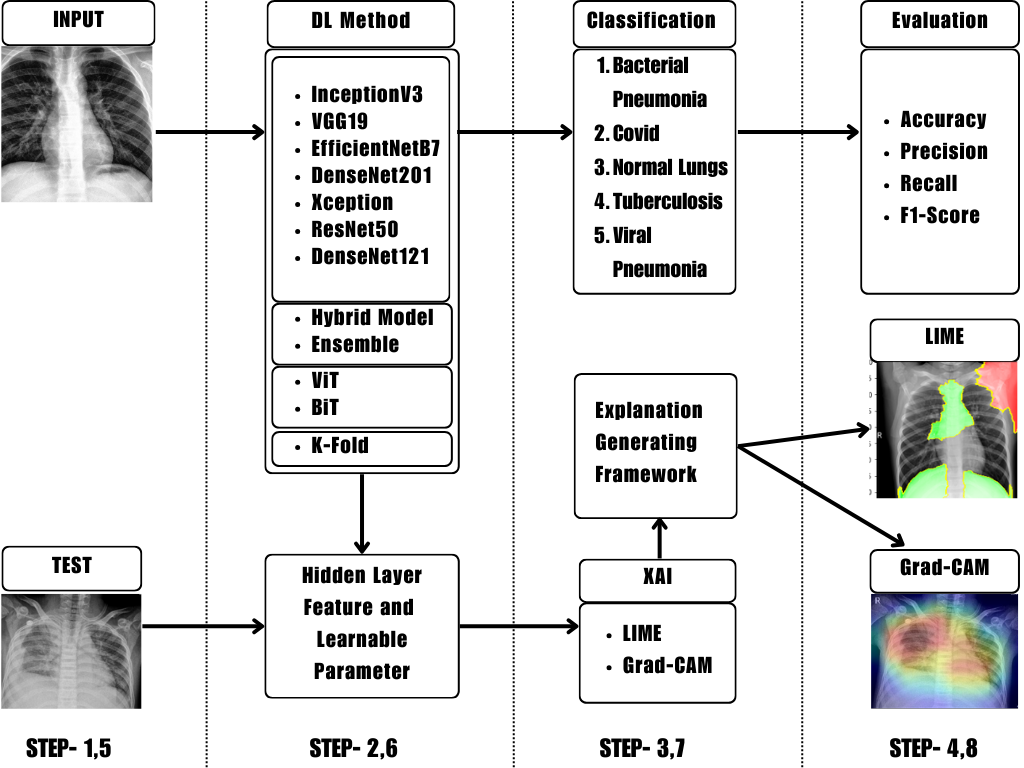}
\caption{This methodology for Lungs Disease identification: The visual feature extractor module is started from the left side blocks}
\label{fig:my_figure}
\end{figure}

The architecture for identifying lung illness from chest X-ray pictures utilizing explainable artificial intelligence and deep learning is shown in this part. Our model has two parts, the first is dedicated to capturing visual attributes, and the second is focused on classifying the images into five classes. We will now provide a breakdown of each step within this architectural framework \newline
 
\textbf{Step-1,5) Input Image:} First of all, we partitioned the dataset, allocating 59.67\% for training, 20.06\% for testing, and an additional 19.97\% for validation. In this stage, we present the proposed model with batches of lung images from the training dataset. For our work, the batch size is 32.\newline
\textbf{a) Image Augmentation:} First, we ensure that all images are the same size and to enable more effective processing, we pre-processed each image by scaling it to 299*299*3. Normalization technique is used to reduce the image pixels will be scaled between 0 and 1. This helps to stabilize the training process by standardizing the pixel values to have an average of 0 and 1 value is for a standard deviation. Also, we use different types of augmentation techniques such as

\textbf{a) Rescaling:}
The parameter \texttt{rescale=1./255} normalizes image values to the range [0, 1], a standard step in image classification preprocessing. This ensures the uniform pixel value ranges, making the data more suitable for neural networks.

\textbf{b) Shifting:}
The values of 0.2 for the width shifting range and height shifting range parameters cause pictures to be randomly shifted both horizontally and vertically. This enhances the diversity of the dataset and helps to create a model that is adaptable to small changes.

\textbf{c) Zooming:}
The \texttt{zoom\_range=0.2} parameter randomly adjusts image zoom by up to 20\%, enhancing the model's ability to handle different sizes and improving its accuracy and generalization with new images.

\textbf{d) Flipping:}
By arbitrarily mirroring images along the horizontal axis, the setting \texttt{horizontal\_flip=True} doubles the dataset with flipped versions, allowing the model to detect features nonetheless of their left-right positioning.

\textbf{e) Rotation:}
The parameter \texttt{rotation\_range=10} presents random rotation of images by up to 10 degrees. This feature enhances the diversity of alignments within the dataset, thus upgrading the model's capacity to adapt and perform accurately with new images.

\textbf{f) Shearing:}
Up to a maximum of 20\%, the setting \texttt{shear\_range=0.2} presents arbitrary shape variations to images through horizontal or vertical contortions. This development strengthens the model's resistance to form changes and enhances its flexibility with new photos.

\textbf{Step-2,6) Deep Learning Methodology:} Utilizing deep learning algorithms, deep learning methodology requires a methodical process for controlling complex circumstances \cite{22}. The stairs involved in this procedure are problem definition, data collection and preprocessing, model construction, training, evaluation, hyperparameter tuning, and model deployment. Deep neural networks are appointed to extract significant information and decipher complex patterns from giant datasets. Through continuous testing, development, and refining, these models are developed to generate predictions on previously unseen data.
\newline
\textbf{a) Visual Feature Extractor:}
Seven well-known models were used in our study to extract visual characteristics from the dataset: Xception, Inception-V3, VGG19, EfficientNetB7, DenseNet201, DenseNet121, and ResNet50 \cite{23}. The training settings comprised 6000 iterations, a number of batch size is 32, and a rate of learning of 0.0001. Each epoch—which was established by the number of iterations needed to cover the whole training dataset once—had a total of 31. Applications for multi-class classification can benefit from the use of the optimizer known as Adam and the function for loss is known as categorical cross-entropy., were used for optimization and loss computation. The neural network's parameters were set by this extensive training setup and subsequently assessed with test data.
\newline
\textbf{a) Hyper-parameter Tuning:} Changing a model's hyper-parameters is essential to increasing its efficacy. The grid search approach was used to find the best combination of settings for this goal. The optimizer named Adam was selected with a learning rate of 1e-4, a dropout rate of 0.6 to avoid overfitting, 32 batch sizes for effective learning, and 31 epochs to maintain a balance between learning and depend on training data.\newline
\textbf{b) Transfer learning and Fine Tune: }Transfer learning is a process that uses past information to modify a model that has been trained on one job for a related one. In neural networks, this frequently means taking a model that has already been trained on a generic task and applying it to a particular, related task. The provided code snippet fine-tunes the last 50 layers of a pre-trained DenseNet121 model for a specific task, allowing adaptation without retraining the entire model. This is done by allowing these layers to be trainable throughout the training, ensuring they can adapt to the specific details of new data.\newline
\textbf{c) Hybrid Model: }We combined InceptionV3, VGG19, and Xception in hybrid models for a multi-class classification task, achieving up to 89\% accuracy. Carefully chosen hyperparameters and data augmentation enriched the training process. Results emphasize the batch size impact on training dynamics, offering insights for further optimizations in model architecture. \newline
\textbf{d) Ensemble Learning: }Ensemble models combine multiple individual models to improve predictions. Common strategies include averaging predictions, assuming equal contribution, and majority voting, relying on the most frequent prediction. This approach explores diverse model architectures for enhanced image classification performance.\newline
\textbf{d) Stratified K-fold cross-validation: }It is tailored for imbalanced datasets, enhancing the evaluation of classification models. It ensures balanced representation in each fold, preventing bias toward the majority class. Each of the k folds (usually k=5 or k=10) in the dataset serves as a testing set for the model as it is trained k times. Average performance across folds estimates the model's generalization error. In the resource problem, two K values, K=3 and K=5 with random\_state=42, are used for dataset folding.\newline
\textbf{e) Vision Transformer and Big Transfer: }ViT and BiT were compared on an image classification task \cite{22}, trained with identical hyper-parameters and duration (79 epochs). ViT outperformed, indicating its superior architecture and training approach. Both models shared the same standardized hyper-parameters for fair comparison, emphasizing the impact of architectural differences on performance.\newline

\textbf{Step-3,7) Classification and Explanation with XAI:} We compile a list of models to run for our visual tasks. Then, we execute all of the visual models with the identical hyper-parameter configuration and save event history. All the values of each epoch is stored in history. So in step 4, the model's predictions among five distinct respiratory disease classes such as Bacterial Pneumonia, Corona Virus Disease, Normal, Tuberculosis, and Viral Pneumonia are subjected to thorough visualization using explain ability techniques.
In step-8 the process involves preparing test images, running model predictions, and applying LIME \cite{8} and Grad-CAM \cite{7} to samples. The resulting visualizations provide interpretable insights into the model's decisions, fostering transparency and trust in critical applications like medical image classification.\newline

\textbf{Step-4,8) Evaluate The Models and Visualization Predictions:} In step-4, confusion matrix is used to compare performance. The model's miss classification rate has been utilized as one of the metrics to effectively compare its performance across several classes. To evaluate how well the model performs, we utilize the weighted F1-score measure. Finally, in step-8 LIME is utilized to generate local explanations for individual predictions, offering insights into the specific features influencing the model's decision-making. Grad-CAM offers an illustrative heat map that highlights key areas within the input photos that are important for the model's classifications.
\section{Result Analysis}

We implemented several models, and the results obtained from those models are shown in Table (3-8). 

\textbf{a) The Result by Pre-trained CNN Model:} We conducted experiments using seven models. The results of a classification test using seven CNN models that have been trained beforehand. At 92\% accuracy, DenseNet121 and InceptionV3 are second and third, respectively, to Xception's \textbf{93\%}. DenseNet201 attains 91\% accuracy, whereas ResNet50, EfficientNetB7, and VGG19 vary in accuracy from 90\% to 86\%. The models exhibit consistent performance, as seen by the tight alignment of precision, recall, and F1 scores with accuracy. Xception performs admirably as the best model for the given job.

\textbf{b) The Result by Hybrid Model:} Hybrid models were created using various combinations. The InceptionV3+Xception hybrid is particularly noteworthy as it attains the greatest accuracy of 89\% while exhibiting constant F1 scores, precision, and recall. With 88\% accuracy, other combinations such as InceptionV3+DenseNet201 and VGG19+DenseNet201 also perform well. These results highlight the value of mixing different CNN architectures and demonstrate how hybrid models may perform better in the given job in terms of prediction.

\textbf{c) The Result by Ensemble Model:} Ensemble models are created through multiple combinations using two criteria: average prediction and majority voting. The highest accuracy of 93\% was achieved, surpassing other combinations. In the "Average Prediction" table, various combinations involving Xception, InceptionV3, DenseNet201, VGG19, and ResNet50 consistently yield an accuracy of 93\%, reflecting high precision, recall, and F1 scores. The "Majority Voting" table follows a similar pattern, with the top accuracy of 93\% obtained through ensemble combinations.

\textbf{d) The Result by Vision Transformer and Big-Transfer Model:} We also implemented Vision Transformer (ViT) and Big-Transfer (BiT) models. Vision Transformer achieved an accuracy of 91\%, while Big-Transfer achieved an accuracy of 86\%.
\begin{table}[h]
 \caption{Result obtained by transformer and big transfer model}
  \label{tab:example}
  \centering
  \begin{tabular}{|c|c|c|c|c|}
    \hline
    \textbf{Model Name} & \textbf{Accuracy} & \textbf{Precision} & \textbf{Recall} & \textbf{F1} \\
    \hline
    Vision Transformer (ViT) & 91 & 91 & 91 & 91 \\
    \hline
    BigTransfer (BiT) & 86 & 86 & 86 & 86 \\
    \hline
  \end{tabular}
\end{table}

\textbf{e) Stratified K-fold Cross Validation}
We implemented the K-fold technique to determine the best accuracy. Utilizing both 3-fold and 5-fold cross-validation, we applied this technique to the base dataset. We utilized the Xception model in this K-fold analysis since it showed the best accuracy in a single model evaluation.

\textbf{I) Using 3-fold:} The model achieved an accuracy of 90\% on one fold, 95\% on another fold, and 98\% on the third fold.
The average accuracy across all three folds is 94.33\%.
Precision, recall, and F1 scores are all very high, ranging from 94\% to 95\%.

\textbf{II) Using 5-fold:}
The model achieved an accuracy of 90\% on one fold, 95\% on another fold, 98\% on another fold, 99\% on another fold, and 99\% on the fifth fold.
The average accuracy across all five folds is 96.20\%.
Precision, recall, and F1 scores are all very high, ranging from 96\% to 97\%.

K-fold cross-validation ensures more efficient use of data and minimizes bias, leading to enhanced accuracy compared to other methodologies. This comprehensive approach to model evaluation significantly contributes to the reliability of our results, providing a robust estimate of the model's generalization capability on unseen data. 
\begin{table}[h]
\caption{Result obtained by Stratified K-fold Cross Validation model}
  \label{tab:examples}
  \centering
  \begin{tabular}{|c|c|c|c|c|c|}
    \hline
    \textbf{Scene} & \textbf{Model} & \textbf{Accuracy} & \textbf{Precision} & \textbf{Recall} & \textbf{F1} \\
    \hline
    3-fold & Xception & 94.33 & 94 & 94 & 94 \\
    \hline
    5-fold & Xception & 96.20 & 96 & 96 & 96\\
    \hline
  \end{tabular}
\end{table}

\section{Conclusion and future work}
The study explores automated diagnosis of lung disorders using Chest X-ray images. It proposes models to classify and analyze lung diseases using XAI tools like the LIME algorithm and Grad-CAM. The Xception model achieved a 96.21\% accuracy using stratified 5-fold cross-validation, demonstrating its usefulness in improving lung disease diagnosis. Future research should explore integrating imaging modalities, hybrid models, image segmentation \cite{23}, and SHAP to enhance accuracy and robustness. Investigating real-time applications and deploying the proposed models in clinical settings could further validate their effort and effectiveness in streamlining the diagnostic process for lung disorders. Furthermore, collaborative efforts between medical professionals and machine learning experts are crucial for refining models and ensuring their practical utility in diverse healthcare scenarios.

\vspace{12pt}
\end{document}